# Audit and change analysis of spreadsheets

John C. Nash, Neil Smith, Andy Adler
School of Management, University of Ottawa
136 J-J Lussier Private, Ottawa, Ontario, K1N 6N5, Canada
jcnash@uottawa.ca

**ABSTRACT**

*Because spreadsheets have a large and growing importance in real-world work, their contents need to be controlled and validated. Generally spreadsheets have been difficult to verify, since data and executable information are stored together. Spreadsheet applications with multiple authors are especially difficult to verify, since controls over access are difficult to enforce. Facing similar problems, traditional software engineering has developed numerous tools and methodologies to control, verify and audit large applications with multiple developers. We present some tools we have developed to enable 1) the audit of selected, filtered, or all changes in a spreadsheet, that is, when a cell was changed, its original and new contents and who made the change, and 2) control of access to the spreadsheet file(s) so that auditing is trustworthy. Our tools apply to OpenOffice.org **calc** spreadsheets, which can generally be exchanged with Microsoft Excel ®.*

## 1. MOTIVATIONS AND BACKGROUND

Scandals and debacles in the business world underline the need for better ways to audit and analyze business data, especially in the form of spreadsheets. Spreadsheet tools and models are used as the basis for many business, administrative and engineering decisions. The decision-makers need to have confidence that their spreadsheets contain reliable information, free from errors or malicious changes. Spreadsheets themselves – apart from the programs that process them -- can be complex and powerful programs, yet seldom are they submitted for quality audits or code assessments, as is standard practice in the software industry. (Fagan, 1976; Ebenau et al., 1994; Gilb and Graham, 1993; Dunsmore et al., 2001; Knight and Myers, 1993; Laitenberger and DeBaud, 2000.) Auditors, as well as spreadsheet designers and owners, need better tools to determine when, why and how an error was introduced to determine its impact on decisions based on the spreadsheet's results. Some such tools were reviewed by Nixon and O'Hara (2001), but to our knowledge none of these can provide a true change-analysis audit trail, though clearly there are many points of similarity with our work.

Current programs such as OpenOffice.org **calc**, Microsoft Excel ® or Corel Quattro ® (hereafter called Calc, Excel and Quattro respectively) have the ability to record and manage changes made to spreadsheet documents. These capabilities are, however, focused on reviewing and perhaps "rolling-back" changes in a chronological fashion. Furthermore, the roll-back, recorded in the "change_history", requires special care to establish, typically has a time limit, and may be lost accidentally or intentionally by user actions. It is generally left to the reviewer to manually determine the potentially damaging change in the list of entries. Our work addresses the need for software tools to manage change control of spreadsheets. We have developed a cross-platform spreadsheet system based on the **calc** file structure. While a large percentage of spreadsheet users work on computers run under variants of Microsoft Windows, we note that there is a growing community of workers employing flavours of Unix or Linux. Our work is intentionally cross-platform; we work on Microsoft, Unix, Linux and Macintosh platforms, and have tried to ensure our results are similarly portable.

A related issue is the existence of computer viruses and similar destructive code that may alter spreadsheet contents. Archives of virus-protection companies list several spreadsheet viruses such as *XM97/Laroux-OC, Excel.Extra, Extra.xls* and *Style.A*. (http://www.sophos.com/virusinfo/, http://securityresponse.symantec.com/avcenter/, http://www.mcafee.com/ ; except where mentioned explicitly, all sites in this paper were visited on 2003-4-10 between 0250 and 0315 GMT).

A motivating application for two of us (JN, AA) is that of maintaining student marks files when there are multiple markers. If the professor keeps a master file and merges partial spreadsheets into it from markers, there inevitably comes the day when a marker sends a second version of a partial file with "sorry, I made a mistake". Un-merging the faulty data is a daunting task. If there were an audit trail, one could rebuild the master file to the point of the merge and thereby undo the damage. (This task is, however, still on our "todo" list.) Constrast these problems with software engineering, where many sophisticated and integrated suites of tools exist for the management and control of large software projects with multiple developers [ SCCS: Bolinger and Bronson, 1995; RCS: http://www.gnu.org/software/rcs/rcs.html; CVS: www.cvshome.org; MKS: http://www.mks.com/products/sie/; Visual Source Safe: http://msdn.microsoft.com/ssafe/; BitKeeper: http://www.bitkeeper.com/, Subversion: http://www.subversion.org (visited 20030616 at 0900 GMT approx.) or Rational Clearcase http://www.rational.com ]. While it is somewhat difficult to compare the size of a spreadsheet with traditional software, a simple measure of counting development hours would place a large spreadsheet on a comparable footing to many commercial software offerings. Support for audit, verification, and change control for spreadsheets is in its infancy compared to traditional software. Spreadsheets, being a mix of code, data and presentation (graphics) elements, are perhaps more difficult to maintain and verify. Change analysis tools like CVS have been designed to work on line by line changes in text documents; as such, they would be difficult to use without suitable formatting or analysis of the results.

One solution to the situation above, of course, is to use a database system to manage the marks. However, there are many situations that arise where special policies must be applied for students who have been excused certain requirements or overall marks must be adjusted in some way. These are more conveniently handled with a spreadsheet. Moreover, the effects of adjustments can be tested before they are committed to file.

Outside the "marks" environment, the building of "private" or "black book" spreadsheets is a practice much denigrated by senior managers and auditors, but because of the awkwardness of use of some centralized systems and the need to test "what if" scenarios privately it seems certain to continue. As a topic for discussion, we contend that it may be better to seek to provide appropriate methods by which "private" spreadsheets (or equivalent functionality) are properly maintained so that they may be easily and controllably integrated with central systems. We believe our work is helpful in this regard.

The popularity of Excel would normally lead us to use Excel spreadsheets as the basis for our work on audit and change analysis. However, as we have noted, it does not (yet) operate on Unix/Linux systems. Moreover, the code and file formats are proprietary, while *calc* has source code available should we need to modify it to control the change analysis. Its file format is built on compressed XML.

## 2. ISSUES IN AUDIT AND CHANGE ANALYSIS

We talk of "Audit and Change Analysis" together because auditing is built upon the analysis of transactions – that is, of changes and relationships in information files. The better integrated this function can be, the more efficient and successful the audit process becomes. Having an easy and efficient method of reviewing changes makes inclusion of change analysis into the design and support of spreadsheets more likely, thereby improving quality and reducing support costs.

In auditing spreadsheets, we need to understand cell relationships. This understanding becomes harder as sheets are added and linked or referenced to each other, and worse still if references are made to external spreadsheets. Generally we will want to determine potential high-risk "zones", that is, particular users, types of cell changes, times of day or month, where we believe or have experience of unwanted changes. To work efficiently, the auditor wants to know where trouble may lie, for example, the 20% of cells that need 80% of the attention. Auditors need to see cell contents, not just displayed results. A formula and a static value can appear the same on a display, but have very different implications. Spreadsheets allow the contents of cells, that is formulas, values, macros, references, etc., to be displayed as well as the results and formatting, and even some indication of other cells involved as in a sum or similar formula. These tools however are relatively limited in utility to the auditor. Knowledge of the exact meaning and ordering of parameters of functions may be critical. Tools can help by providing support to relieve the "mechanical" burden of such syntactic knowledge, allowing the auditor to focus on the impact of the relationships between spreadsheet elements. Similarly, a tool may be used to highlight particular situations of interest to the auditor.

Static analysis of spreadsheets looks at a single instance of the spreadsheet to identify errors. We may want to examine address references to see if there are copy or fill errors (i.e., a reference does or does not change with respect to the change in the referring cell), blank cells (which could be user input, however if the referring cell does not check for absence of an entry, other results could be invalid), or ranges (where insertions or deletions at boundaries of the range may not be updated in referring cells). Built-in functions can also be examined, for instance, to see if parameters are assigned correctly, have the right type (cell referencing versus use of constants as in the NPV function: where one may want to use a cell reference to allow for a changeable interest rate, but a user may "temporarily" set it to a fixed value), or even to review the type of functions used. For example, is the existence of trigonometry functions in a sales forecasting spreadsheet cause for concern?

Static analysis may also pick up what we call "constant equations". For example, a cell may have the contents " =1+2+3". This is not invalid, but we may question its purpose. (Change analysis may give a clue.). Errors such as reference errors, invalid type and other errors flagged by the spreadsheet program during calculations may also be picked up. While they should be obvious, since the spreadsheet program will flag them, they may not be easily visible in a large spreadsheet. Name references present the same issues as address references, plus others. Duplicate references to same cell are valid, but could be an error or an opportunity for simplification. Referenced address ranges need checking for correctness, in particular for overlapping ranges?

Change Analysis of spreadsheets also presents difficulties. In the life of a spreadsheet, most changes are unlikely to cause errors, and "Checkpointing" or "versioning" of the spreadsheet files helps to decompose our problem. That is, once an auditor is satisfied with the spreadsheet up to a given checkpoint, effort can be focussed on changes made after that version. This also can assist in reducing the volume of change information that must be stored. On the other hand, spreadsheets can have many changes made by many users. While plain (static) data may have errors, these are relatively simple to correct. It is the executable cells that may be much more dangerous since their output affects other cells, but such errors may be largely invisible. Auditors will likely want to focus on the "program" rather than various "executions". Correct use of cell or sheet protection helps to ensure the integrity of the spreadsheet. A final woe: recording changes is optional with all spreadsheet programs familiar to us – we have to ensure it is turned on to create our audit trail and protect the integrity of this trail.

Dynamic or change analysis implies that we must look at change information captured during spreadsheet editing sessions. When coupled with static analysis this adds the dimension of time, which allows the user or auditor to determine the impact of an error on decisions based on the spreadsheet. Analysis of changes covers a number of segments in a spreadsheet. First, administrative information such as the author / user making the change, and the date or status of the change. We will want to be able to focus attention on specific changes, such as the type of change (content, insertion or deletion), or the change action. Of particular interest will be dynamic to static changes, such as

- Cell or name reference changed to static value. This could be either a single cell reference, or as part of an equation. For example, if a particular cell represents a currency exchange value or discount rate, users may expect this to be used for all relevant calculations, so removal of references may invalidate this assumption.
- Built-in function replaced by a static value. As educators, we see immediate possibilities in detecting potential cheating where cell results are set as constants when they should be computed or derived, or where computed marks are replaced with (usually higher) numbers.
- Built-in function parameter changes. Cell or name references changed to static values. We should be concerned with the influence on "what-if" scenarios when a referenced or computed value is not used.
- Function or name reference changes.
- Built-in function replacement. Example: PV(…) changed to FV(…); same parameters, different results.
- Name reference changes. Example: Profit_2002 changed to Profit_1999. This is a "documentation" change.
- Cell or sheet protection. This is important and necessary to ensure the integrity of the spreadsheet when used by clients. Design changes require the removal of protection and therefore opportunity to not restore protection before delivery to client. A change to sheet protection leaving all cells unprotected may be very dangerous, for example, in a spreadsheet delivered to a client for completion and return (tax calculation, sales forecast), which could be accidentally or intentionally changed and not detected when returned. Cell protection status also needs to be verified, since accidental or intentional changes to individual cell protection can provide integrity "holes" in spreadsheets used by clients.

We believe spreadsheet change analysis has applications in all areas where spreadsheets are used. While we have described this in terms of "auditing", users can employ the tools we describe to better debug complex spreadsheet packages.

Aware of these issues and the "wish list" of items, the authors developed a program, SSScan, that combines static and change analysis. The current version of SSScan incorporates some of the items listed above.

**3. THE SSSCAN SPREADSHEET AUDIT TOOL**

*SSScan* is a tool being developed to combine spreadsheet content analysis and change record history, thereby addressing most of the issues in the previous section. *SSScan* analyses *calc* spreadsheets containing change information and directs the user to suspicious changes. By simply configuring various filter settings the entire document (containing single or multiple sheets) can be quickly and easily checked for changes of the specified type. SSScan also provides a visual display of the cell contents and inter-relationships. It functions as a stand alone, read-only program to enhance audit process audit ability. Note that *calc* is able to read and write almost all Excel spreadsheet files. Spreadsheet auditing software exists for Excel files (ref Nixon 2001, details on EUSPRIG site). We know of no such software for *calc* spreadsheet files.

As mentioned, while tools exist for assisting spreadsheet review (Nixon and O'Hara, 2001), to our knowledge none but SSScan can create a complete audit trail. That is, we want to know that a cell was changed from X to Y at time T by user U, and be able to reconstruct the spreadsheet at any stage. Clearly, for most applications, one is likely to choose a less detailed view. However, when there is a serious concern, SSScan can make available the fundamental data. (To date, we have not built a tool to reconstruct the spreadsheet up to the point of a particular change.)

Current types of filters include the standard:
- Author name
- Date range
- Cell address range
- Cell content

Stronger analysis is possible by cell content filtering. Examples of settings include:
- Formula cells changed to static cells, for example, changing "=Sum(…)" to "=100" to *force* the desired result
- Initial cell entries (that is, entering information into a blank cell)

All filters can be set to be inclusive or exclusive and wild cards are allowed. Additionally in the case of Author name, case can be ignored. Finally, multiple filters can be enabled to allow for a complex analysis. For example:
1. select all changes made by "J* Doe",
2. between "December 24, 2001 and January 1, 2002", that
3. were not initial entries.

In setting (1), "*"finds "J. Doe, Jane Doe, John Doe or other combinations starting with "J" and ending with "Doe". The user could also set "case ignore" to *true* to capture further combinations that might have been missed.

*SSScan* is written in Java and the same binary file installs on Windows, UNIX/Linux and MacOS platforms. Currently only *calc* files (.sxc files) can be analysed. Excel can be imported by *calc* and then saved as .sxc format for successful processing by SSScan.

*SSScan* is a stand-alone program that only reads a spreadsheet file. We believe that this is helpful to auditors since there is no possibility of corrupting the data within the spreadsheet file. Many programs that deal with spreadsheets are "add-ins" to the relevant spreadsheet program (Excel®, Quattro®, Lotus ®, etc.) Thus it is not possible to state with certainty that the auditing activity has not corrupted the original data.

Figure 1. SSScan change record panel showing cell content changes.

Figure 2. SSScan filter panel.

## 4. PROTECTING SPREADSHEET FILES

A serious problem with any spreadsheet file, as illustrated by the marks-merge example above, is that spreadsheets usually exist on the user's current machine. In many situations, the user's name is not verified or controlled. Version and authorship control are therefore issues if we are concerned about errors. Malicious changes are a worry if we have content that is sensitive. Moreover, if a malicious user has time and knowledge, he/she can make changes and also cover the evidence of those changes. Our audit tool cannot find what has been carefully deleted with such diligence in hiding the trail of actions.

Exceland Quattro both have capabilities for sharing notebooks and even for reconciling conflicts when two or more users open a file at the same time. (Clearly, the assumption is that the file is accessible to more than one user at a time on some sort of network storage.) We feel a better solution to this issue is to keep the spreadsheet file in one location where it can be protected by access controls. Furthermore, it will be processed by a spreadsheet program at that protected location. This is possible using client-server technology and the Internet. Users access the spreadsheet files that reside on a server from their local workstation over a network. A spreadsheet program on the server is used to manipulate the spreadsheet, and access controls limit manipulation of the spreadsheet to this program. Figure 3 illustrates the flow of information.

There are a number of technical issues and options in implementing the processes in Figure 1. In particular, modern spreadsheets are essentially graphical in nature, so we need to present a screen as if it exists on the local workstation of the user. Fortunately there are tools that allow this, though they are not trivial to configure and use. Moreover, they often require high communications bandwidth to give a satisfactory user experience.

Figure 3. Diagram of user access to spreadsheet file ss1.

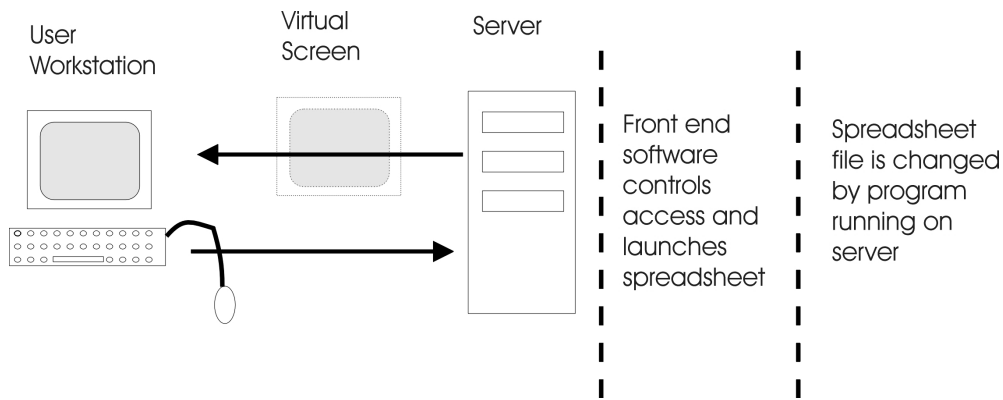

The essence of such tools is to extract from the server's screen image appropriate information that allows the "drawing" of this image on a remote, client screen. This remote machine may have different properties from the original and may be run by a different operating system. Within the Microsoft DOS/Windows environment, many users are familiar with PC Anywhere as such a tool. For those running Unix/Linux servers, there is the possibility of running remote X11-server applications (http://www.linux.org/docs/ldp/howto/mini/Remote-X-Apps.html). (X11 is the name of the umbrella software that draws the screens on Unix machines.) To access a Unix or Linux server in this way requires that the Windows client computer has suitable software such as ***X-win32*** (http://www.starnet.com/) or ***MI/X***        4

(http://www.microimages.com/mix/). We actually got the *gnumeric* spreadsheet running on a Linux machine but viewable on a Windows client machine in this way with the free trial of *MI/X*. However, there appear to be no reliable open source or free tools for remote X and the protocols are quite demanding of communications bandwidth. A related approach is VNC (for Virtual Network Computing) developed by a community of workers (see http://www.uk.research.att.com/vnc/). Moreover, there are two quite reasonable implementations that are freely available, namely RealVNC (http://www.realvnc.com) or TightVNC (http://www.tightvnc.com) There is also Microsoft Terminal Services Advanced Client http://www.thin-world.com/tsweb/readme.htm, but we did not investigate this in detail. Indeed, we chose to work with VNC (Virtual Network Computing).

Another important consideration is security. Network interfaces to graphical communications should be encrypted to prevent eavesdropping and injection of false data. While all network graphical interfaces can be cryptographically secured, VNC is the most flexible and easy to secure. All VNC communications take place over a single TCP/IP socket connection, which can be protected using Secure Shell (SSH:, (See http://www.ssh.com or http://www.openssh.com ) Secure Sockets Layer (SSL: http://www.openssl.org) or IPSec tunnels (http://www.linuxsecurity.com/docs/LDP/VPN-Masquerade-HOWTO.html) using standard tools.

We also chose to host our spreadsheet files and spreadsheet program on a Linux server, mainly because we believe that there is more flexibility in configuring such servers. However, we continue to experiment with Windows and Macintosh implementations in the spirit of platform independence already mentioned. Regardless of our platform, we need to carry out the following operations in the correct sequence:

- Allow a user U to "log in" to the server in some way that only presents U with a screen of files that U is allowed to access. We have implemented this as a Web-based login screen written in Perl. The users have no accounts on the server, but are given usernames and passwords within the Perl login scripts. Such users have no privileges on the server other than those granted through the Web pages.
- Colour-code links to the permitted files are presented, showing files already in use as read-only (e.g., in red) or read-write (e.g. in green). That is, we chose to carry out *file locking* within our scripts. We have not yet implemented this feature, but the programming required is straightforward.
- On selection (clicking) on a particular file, launch the spreadsheet program to access ONLY the selected file, and to pass the username U to the spreadsheet program. The filename is passed by putting its name on the command line after /usr/local/bin/scalc for example. ?? verify internal name of calc We have yet to restrict the File/Open and File/SaveAs dialogues of *calc* that could possibly be used to access other spreadsheet files. There are several potential mechanisms to control such access. The username is passed by modifying a file called UserProfile.xml that is stored in the OpenOffice program folder.) We must also ensure that the Change Recording is turned on. This is controlled by a setting within the spreadsheet file. Note that the launch of *calc* must also invoke vncserver. We may need to arrange that the proper password for user U is reset into vncserver. We also need to ensure a vncviewer is active on the client workstation, but there is a Java version of vncviewer that is usable as an applet in a Web browser. One of the advantages of the VNC framework is that no software installation is required on the client workstation. A client viewer is automatically downloaded and executes in the JAVA virtual machine in the client browser. Another advantage of this framework is the easy integration with active web server content which allows browsing and selection of available files.

- We must allow U to save the file; automatically keeping the audit trail. We intend to add the capability of backing up the spreadsheet with checkpoints, but to date have not implemented this function.
- Finally, we must appropriately terminate the session with the user U so there are no security holes for the next user of the workstation to exploit.

There are other file-locking options open to us to arrange that *calc* can only access files U is authorized to view. *calc* nominally has file locking built in, but we have not tested it to the time of writing this article. It seems likely that a web-based approach will be less problematic.

A final point worth mentioning is that some file formats purport to allow the transfer of spreadsheet information between software packages. SYLK or DIF (http://www.wartburg.edu/compserv/convert.html visited 2003-4-11 at 0100 GMT)files do not, however, appear to support change analysis. It is clear that from the perspective of transparency and comparability of spreadsheets, it would be helpful to be able to audit files in a generic, archivable format. We welcome discussion of this and other issues raised in the paper, with the view to further enhancing the community of expertise in auditing and improving spreadsheets and their usage.

## 6. CONCLUSION AND PROGNOSIS

SSScan is reasonably mature – it can be used now for worthwhile analyses, though there are many more features that we would like to incorporate, for instance the capability of saving a "changes" file that allows stepwise rebuilding of the spreadsheet to a given checkpoint. Providing secure access to spreadsheets so that a verified username and change recording are assured has been demonstrated in its components. That is, we are at the stage of "demonstrated capability" though we not yet a "product" or "system". Our intent is to implement a prototype system by September 2003 so that we can use it to control files of student marks across multi-section courses.

Clearly audit and change analysis for spreadsheets is not a simple topic. If it were, more tools and systems would be available. We believe tools such as those we have described above are important in rendering trustworthy the information that is derived from the use of spreadsheets

## 5. ACKNOWLEDGEMENTS

We are grateful for pointers to some useful references from Prof. Tim Lethbridge of the University of Ottawa. The Ottawa Perl-Mongers, the Ottawa GOSLING Group and the Internal Audit Managers' Group of the Government of Canada have provided us with helpful discussions and brought together the authors. A referee provided some helpful and constructive comments.

# 6. REFERENCES

Bolinger, Don and Bronson, Tan (1995) *Applying RCS and SCCS: From Source Control to Project Control*, Sebastopol, CA: O'Reilly & Associates

Dunsmore, A. Roper, M. and Wood, M. (2001) "Systematic Object-Oriented Inspection: An Empirical Study", Proceedings of the 23$^{rd}$ International Conference on Software Engineering, pages 135-144, May.

Ebenau, R. G. and Strauss, S. H. (1994) *Software Inspection Process*, New York: McGraw-Hill.

Fagan, M. E. (1976) "Design and Code Inspections to Reduce Errors in Program Development", IBM System Journal, vol. 15, no. 3, pages182-211.

Gilb, T. and Graham, D. (1993) *Software Inspections*, Boston: Addison-Wesley.

Knight, J. C. and Myers, A. E. (1993) "An Improved Inspection Technique", Communications of ACM, vol. 36, no. 11, pages 50-69.

Laitenberger, O. and DeBaud, J-M. (2000) "An Encompassing Life Cycle Centric Survey of Software Inspection", Journal of Systems and Software, vol. 50, no. 1, pages 5-31.

Nixon, David and O'Hara, Mike (2001) Spreadsheet Auditing Software, EUSPRIG Conference, 2001, http://www.gre.ac.uk/~cd02/EUSPRIG/2001/Nixon_2001.htm.